\documentclass[%
reprint,
superscriptaddress,
longbibliography,
showpacs,
nobibnotes,
amsmath,amssymb,
]{revtex4-1}
\usepackage[normalem]{ulem}
\usepackage{color}
\usepackage{graphicx}
\usepackage{amsthm}
\usepackage{amsmath}	
 \usepackage{mathtools}
\usepackage{upgreek}
\usepackage{color}
\usepackage{ulem}

\def\U#1{{%
\def\O{\mbox{O}}
\def\u{\mbox{u}}
\mathcode`\u=\upmu
\mathcode`\O=\Upomega
\mathrm{#1}}}
\def\sub#1{_\textnormal{#1}}
\def\sur#1{^\textnormal{#1}}
\def\vct#1{\mathbf{#1}}
\def\cc{\textnormal{c.c.}}
\def\ii{{\mathrm{i}}}

\begin{document}
\title{Anisotropic Babinet-invertible metasurfaces to
realize transmission--reflection switching for orthogonal polarizations
of light} 

\author{Yosuke~Nakata}
\email{y\_nakata@shinshu-u.ac.jp}
\affiliation{Center for Energy and Environmental Science, Shinshu University, 4-17-1 Wakasato, Nagano 380-8553, Japan}
\author{Yoshiro~Urade}
\affiliation{Department of Electronic Science and Engineering, Kyoto University, Kyoto 615-8510, Japan}
\author{Kunio~Okimura}
\affiliation{School of Engineering, Tokai University, 4-1-1 Kitakaname, Hiratsuka, Kanagawa 259-1292, Japan}
\author{Toshihiro~Nakanishi}
\affiliation{Department of Electronic Science and Engineering, Kyoto University, Kyoto 615-8510, Japan}
\author{Fumiaki~Miyamaru}
\affiliation{Center for Energy and Environmental Science, Shinshu University, 4-17-1 Wakasato, Nagano 380-8553, Japan}
\affiliation{Department of Physics, Faculty of Science, Shinshu University, 3-1-1 Asahi, Matsumoto, Nagano 390-8621, Japan}
\author{Mitsuo~Wada~Takeda}
\affiliation{Department of Physics, Faculty of Science, Shinshu University, 3-1-1 Asahi, Matsumoto, Nagano 390-8621, Japan}
\author{Masao~Kitano}
\affiliation{Department of Electronic Science and Engineering, Kyoto University, Kyoto 615-8510, Japan}

\date{Compiled \today }


\begin{abstract}
The electromagnetic properties of an extremely thin metallic
 checkerboard drastically change from resonant reflection (transmission)
 to resonant transmission (reflection) when the local electrical
 conductivity at the interconnection points of the checkerboard is
 switched.
To date, such critical transitions of
metasurfaces have been applied only when they have 4-fold rotational symmetry,
 and their application to polarization control, which requires anisotropy,
 has been unexplored.
 To overcome this applicability limitation and open up new pathways for dynamic
 deep-subwavelength
polarization control by utilizing critical transitions of checkerboard-like metasurfaces, 
we introduce a universal class of
anisotropic Babinet-invertible metasurfaces enabling transmission--reflection switching for each orthogonally polarized wave.
As an application of anisotropic Babinet-invertible metasurfaces,
 we experimentally realize
a reconfigurable terahertz polarizer whose transmitting axis can be dynamically rotated by $90^\circ$.
\end{abstract}

\maketitle

\section{Introduction}
Metamaterials, which are artificial materials composed of engineered
structures, exhibit exotic functionality not readily observed in nature
\cite{Solymar2009}. Two-dimensional metamaterials with
thicknesses much thinner than
the wavelength of light, are called
metasurfaces \cite{Kildishev2013}. The careful design of metasurfaces
enables us to control both the amplitude and phase of electromagnetic
waves \cite{Monticone2013,Pfeiffer2013}. Due to the scale invariance of
Maxwell equations, the extraordinary properties of metamaterials and
metasurfaces can be theoretically realized for all frequency ranges by
altering the size of the structures. In the terahertz frequency range,
metamaterials and metasurfaces are considered promising candidates for
manipulating terahertz waves, because conventional electronics and
photonics technologies cannot be directly applied. Various passive and
active devices based on metamaterials and metasurfaces have been
demonstrated in the terahertz frequency range \cite{Tao2011}.

\begin{figure}[!b]
\centering
\includegraphics[width=\linewidth]{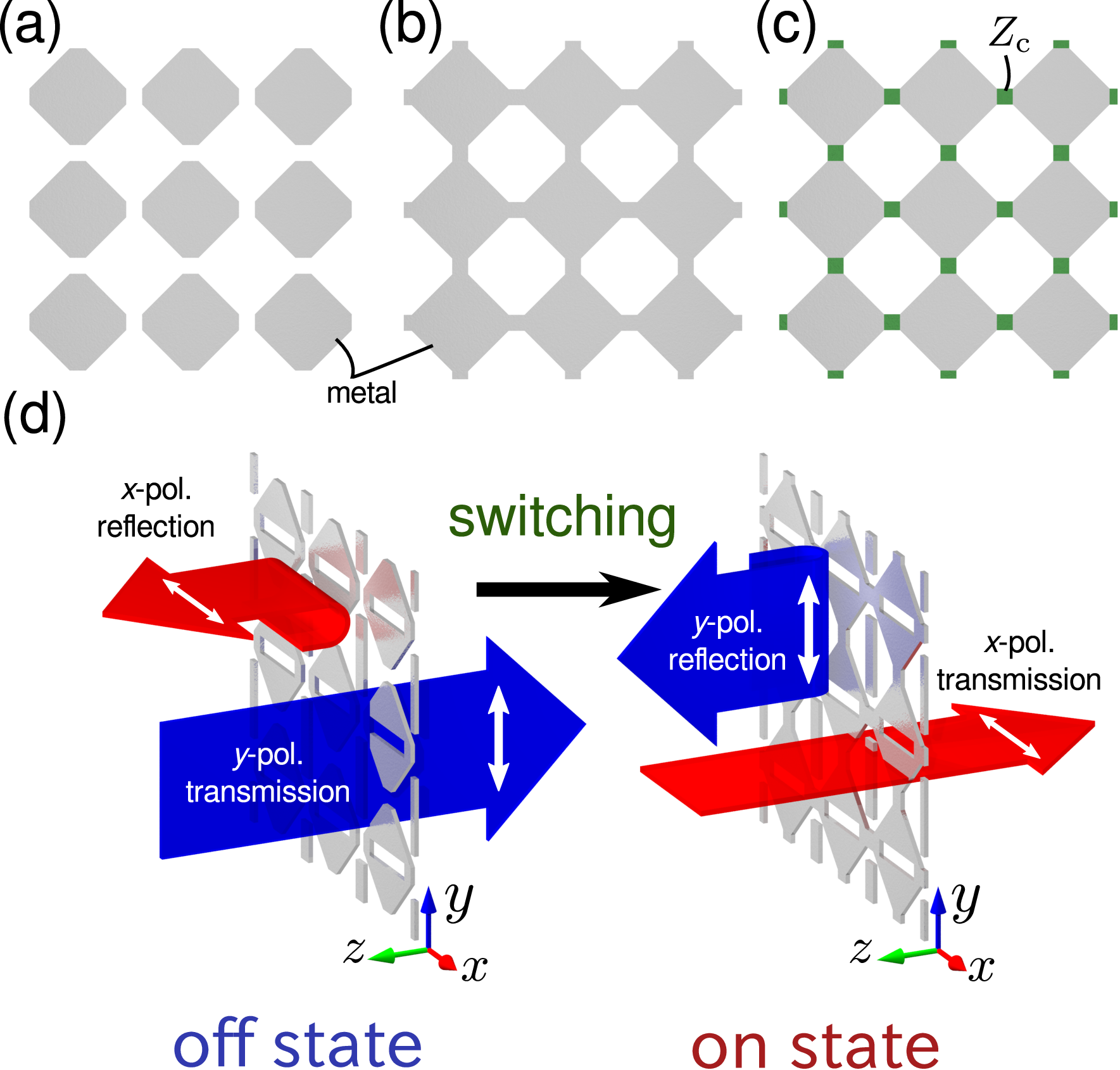}
\caption{(a) Off-state checkerboard metasurface. (b) On-state checkerboard metasurface. (c) Babinet-invertible checkerboard metasurface with variable-resistance sheet with sheet impedance $Z\sub{c}$.
 (d) Dynamic rotation of the transmitting axis of reconfigurable polarizer.
The anisotropic checkerboard can be utilized to realize this functionality.}
\label{fig:checkerboard}
\end{figure}

When designing metasurfaces in a vacuum, Babinet's principle is often
utilized. Babinet's principle relates the scattering fields of a
metasurface to a complementary metasurface. The complementary
metasurface is obtained by applying a structural inversion to
interchange the holes and metallic portions of the given one-layer
metallic metasurface \cite{Booker1946}.
For a more general case, consider
a metasurface at $z=0$
with a spatially varying sheet impedance
of $Z\sub{s}(x,y)$, which relates 
the tangential component (namely, $xy$-component)
of the electric field at $(x,y)$ on $z=0$
and the surface current density at the same point
\footnote{Rigorously, the sheet impedance is defined as
$\tilde{\vct{E}}_\parallel(x,y)=Z\sub{s}(x,y) \tilde{\vct{K}}(x,y)$,
for the tangential component of the electric field
$\tilde{\vct{E}}_\parallel(x,y)\exp(-\ii\omega t)+\cc$ on $z=0$
and 
surface current density $\tilde{\vct{K}}(x,y)\exp(-\ii\omega t)+\cc$ on $z=0$.
Here, $\omega$ is an angular frequency and ``$\cc$'' represents the complex conjugate operation.}. The
complementary surface with sheet impedance
$Z\sub{s}\sur{(comp)}(x,y)$ is obtained by applying a impedance
inversion defined as
$Z\sub{s}\sur{(comp)}(x,y)={Z_0}^2/\big(4Z\sub{s}(x,y)\big)$, where
$Z_0\sim 377\,\Omega$ is the impedance of the vacuum \cite{Baum1974}.

When considering Babinet's principle, the metallic checkerboard 
is a special system: the disconnected [off-state;
Fig.~\ref{fig:checkerboard}(a)] and connected [on-state;
Fig.~\ref{fig:checkerboard}(b)] checkerboards are complementary with
each other \cite{Compton1984}. Their electromagnetic scattering
properties drastically change when the connectivity of the checkerboard
is altered. Such a critical behavior of checkerboard metasurfaces has
been theoretically investigated from the perspective of percolation
theory \cite{Kempa2010} and experimentally observed in the microwave
\cite{Edmunds2010} and terahertz frequency ranges \cite{Compton1984,
Takano2014}. The checkerboard metasurface has also been characterized in
the optical region \cite{Ramakrishna2011}. The critical property of
checkerboard metasurfaces has been applied to reconfigurable
transmission lines \cite{Gonzalez-Ovejero2014a,Gonzalez-Ovejero2014} and
recently the critical behavior caused by rotational disorder has also
been investigated \cite{Tremain2015}.

To control the criticality of the checkerboard, resistive sheets with
sheet impedance $Z\sub{c}$ can be introduced at the interconnection
points of the checkerboard, as shown in Fig.~\ref{fig:checkerboard}(c)
\cite{Nakata2013, Urade2015, Urade2016a}. The reconfigurable checkerboard is a
checkerboard metasurface with dynamically controllable $Z\sub{c}$. If a
metasurface is dynamically switchable to the metasurface congruent to
the complement of the original, like the reconfigurable checkerboard, it
is called {\it Babinet invertible}. Babinet-invertible metasurfaces are
considered to be a generalization of the reconfigurable
checkerboard. The 4-fold rotational symmetry of the reconfigurable
checkerboard and Babinet's principle lead to the relation
\begin{equation}
 \tilde{t}\sur{(off)}(\omega) + \tilde{t}\sur{(on)}(\omega) =1,  \label{eq:1}
\end{equation}
where the (zeroth-order)
complex amplitude transmission coefficients
of
a normally incident plane wave with an angular frequency $\omega$ are
denoted by $\tilde{t}\sur{(off)}(\omega)$ and
$\tilde{t}\sur{(on)}(\omega)$ for the off-state ($Z\sub{c}=\infty$) and
on-state ($Z\sub{c}=0$) metasurfaces, respectively \footnote{Note that
the complex amplitude transmission
coefficient $\tilde{t}$ from an incident wave
$\check{\vct{E}}\sub{0}\exp[\ii (\vct{k}\sub{0}\cdot
\vct{x}-\omega t)]+\cc$ to
a transmitted wave $\check{\vct{E}}\sub{t}\exp[\ii (\vct{k}\sub{t}\cdot
\vct{x}-\omega t)]+\cc$
is defined by $\check{\vct{E}}\sub{t}=\tilde{t}\check{\vct{E}}\sub{0}$.
In this paper, the modifier {\it zeroth-order} represents
the case that $\vct{k}\sub{t}$ satisfies
$\mathcal{P} \vct{k}\sub{t}=\mathcal{P} \vct{k}\sub{0}$, 
where $\mathcal{P}: (v_x,v_y,v_z)^\mathrm{T}\mapsto (v_x,v_y,0)^\mathrm{T}$
is the projection operator onto $z=0$ ($\mathrm{T}$: transpose).}.
Then, the complementary switching of the reconfigurable checkerboard realizes {\it
 transmission inversion}, represented by Eq.~(\ref{eq:1}). The transmission
inversion of the reconfigurable checkerboard has been applied to a
reconfigurable capacitive--inductive terahertz filter \cite{Urade2016}.

 In previous studies of Babinet-invertible metasurfaces, 4-fold symmetry
is assumed and the polarization dependence of the electromagnetic
properties vanishes for normal incidence. Thus, the possibility of
dynamic  polarization control, which is essential for
polarization-sensitive spectroscopy \cite{VanderValk2005,Katletz2012}, 
through complementary switching
 has been unexplored.
In this paper, 
the applicability limitation of the critical transition
of checkerboard metasurfaces to dynamic deep-subwavelength polarization control
is overcome by introducing 
a universal class of
anisotropic Babinet-invertible metasurfaces enabling transmission inversions for each orthogonally polarized wave.
To demonstrate one potential application of the proposed class of metasurfaces,
we experimentally realize a reconfigurable terahertz polarizer as an
anisotropic Babinet-invertible metasurface.

\section{Theory}
If we break the rotational symmetry of Babinet-invertible
metasurfaces, the (zeroth-order) complex amplitude transmission
coefficients $\tilde{t}_x$ and $\tilde{t}_y$ for $x$- and $y$-polarized
incident waves must be distinguished.
Here, we consider an extension of
Eq.~(\ref{eq:1}) to such an anisotropic metasurface as follows:
\begin{align}
\tilde{t}\sur{(off)}_{x}(\omega)+ \tilde{t}\sur{(on)}_{x}(\omega) &=1,  \label{eq:2}\\
\tilde{t}\sur{(off)}_{y}(\omega)+ \tilde{t}\sur{(on)}_{y}(\omega)&=1.  \label{eq:3}
\end{align}
Equations~(\ref{eq:2}) and (\ref{eq:3}) express transmission inversion
separately for the two axes $x$ and $y$. 
We therefore name this
inversion {\it transmission inversion for each orthogonal polarization}, which requires a
specific symmetry.
Note that transmission inversion for each orthogonal polarization
and the electric field continuation condition
$1+\tilde{r}\sur{(off)}_i(\omega)=\tilde{t}\sur{(off)}_i(\omega)$
lead to 
{\it transmission--reflection switching for each orthogonal polarization} represented by
 \begin{equation}
  \tilde{t}_i\sur{(on)}(\omega)=-\tilde{r}_i\sur{(off)}(\omega),  \label{eq:4}
 \end{equation}
where $\tilde{r}\sur{(off)}_i(\omega)$ is a complex amplitude
reflection coefficient for an $i$-polarized normally incident plane
wave.

Next, we discuss the sufficient symmetry to realize
transmission inversion for each orthogonal polarization. 
The rigorous vector version of Babinet's principle
is expressed as follows \cite{Nakata2013}:
\begin{align}
 \tilde{t}\sur{(off)}_{x}(\omega)+\tilde{t}\sur{(comp)}_{y}(\omega)&=1,  \label{eq:5}\\
 \tilde{t}\sur{(off)}_{y}(\omega)+\tilde{t}\sur{(comp)}_{x}(\omega)&=1,  \label{eq:6}
\end{align}
where $\tilde{t}\sur{(comp)}_{i}(\omega)$ represents the (zeroth-order)
complex amplitude transmission coefficient for a normally incident plane
wave with $i$-polarization onto the metasurface complementary to the
off-state one ($i=x,y$) . Note that the polarization is
rotated by 90$^\circ$ in the complementary problem compared with the
original. Comparing Eqs.~(\ref{eq:5})--(\ref{eq:6}) with
Eqs.~(\ref{eq:2})--(\ref{eq:3}), transmission inversion for each orthogonal polarization
requires
$\tilde{t}\sur{(on)}_{x}(\omega)=\tilde{t}\sur{(comp)}_{y}(\omega)$ and
$\tilde{t}\sur{(on)}_{y}(\omega)=\tilde{t}\sur{(comp)}_{x}(\omega)$. These
equations are satisfied when the metasurface complementary to the
off-state surface is obtained by 90$^\circ$ rotation of the on-state
surface. This is the sufficient symmetry condition 
to realize transmission inversion for each orthogonal polarization.

\begin{figure*}[!t]
\centering
\includegraphics[width=\linewidth]{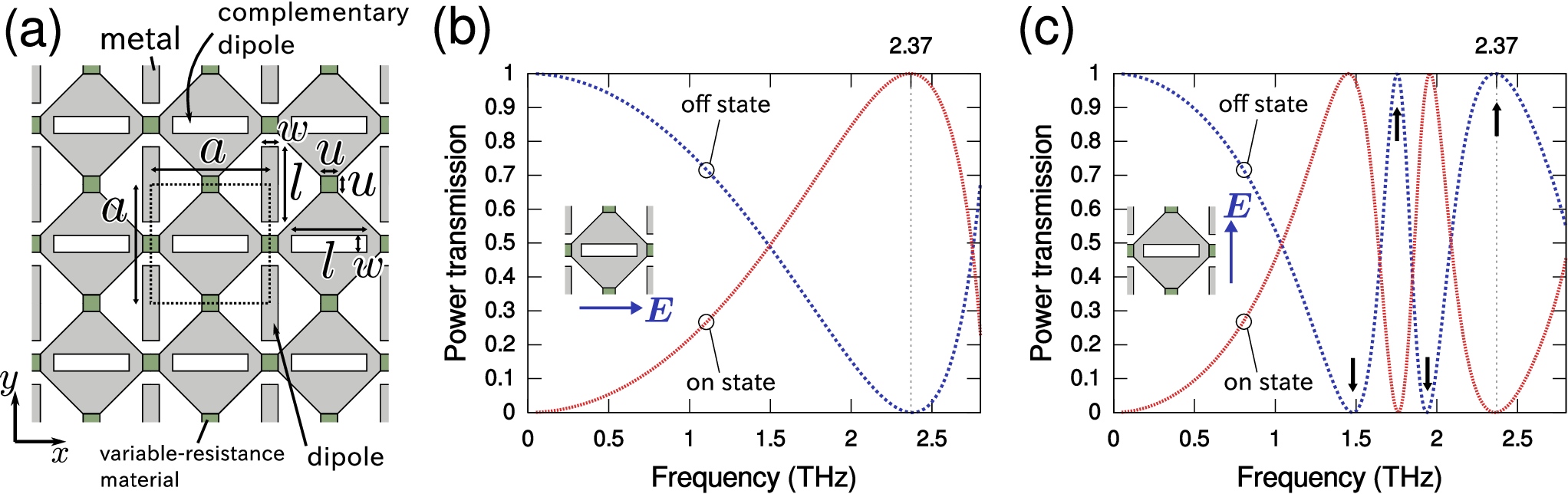}
\caption{ (a) Design of dipole-nested Babinet-invertible checkerboard metasurface (D-checkerboard) with $a=106\,\U{um}$, $u=w=15\,\U{um}$, and $l=67\,\U{um}$. Simulated power transmission spectra of ideal on- and off-state D-checkerboards at the normal incidence of (b) $x$- and (c) $y$-polarized plane waves. }
\label{fig:switchable_d-checkerboard_and_transmission}
\end{figure*}

As an application of transmission inversion for each orthogonal polarization, we consider the
design of a polarizer whose transmitting axis can be dynamically rotated
by $90^\circ$, as shown in
Fig.~\ref{fig:checkerboard}(d).
Designing
such a reconfigurable polarizer with variable-resistance elements
requires simultaneous fine tuning for both the on- and off-state
metasurfaces. This is difficult because the on- and off-state
metasurfaces generally depend on each other. However, by utilizing
transmission inversion for each orthogonal polarization, the problem can be drastically
simplified. If we have an off-state Babinet-invertible metasurface with
$\tilde{t}\sur{(off)}_x(\omega_0)\sim 0$ and
$\tilde{t}\sur{(off)}_y(\omega_0)\sim 1$ at an angular frequency
$\omega_0$, then $\tilde{t}\sur{(on)}_x(\omega_0)\sim 1$ and
$\tilde{t}\sur{(on)}_y(\omega_0)\sim 0$ are automatically satisfied for
the on-state metasurface under transmission inversion for each orthogonal polarization. Then,
the transmitting axis of the polarizer can be dynamically rotated by
90$^\circ$. The problem is reduced to designing the off-state
Babinet-invertible metasurface with
$\tilde{t}\sur{(off)}_x(\omega_0)\sim 0$ and
$\tilde{t}\sur{(off)}_y(\omega_0)\sim 1$.

\begin{figure}[!t]
\centering
\includegraphics[width=\linewidth]{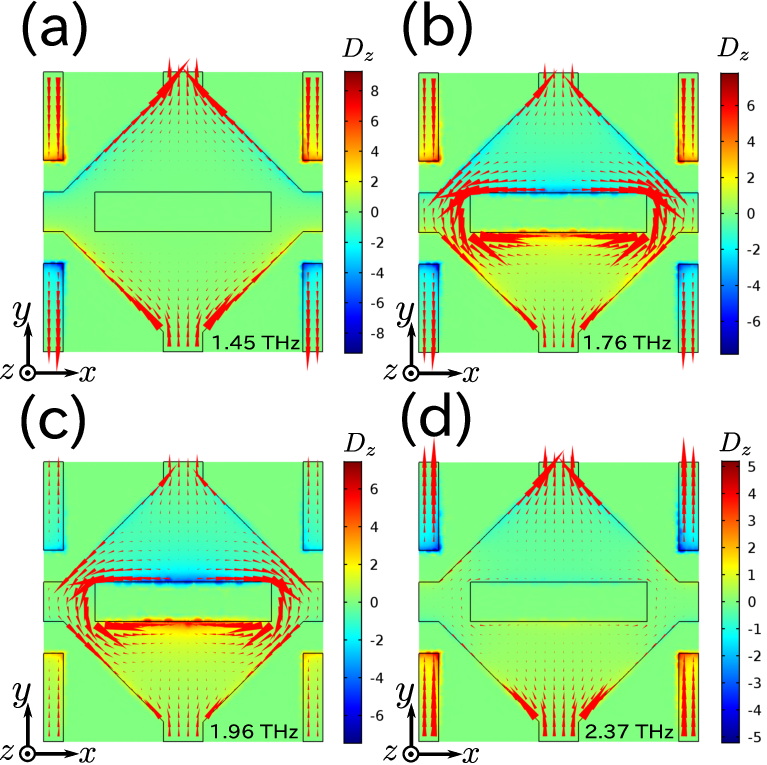}
\caption{Electric displacement $D_z$ on the plane $z=a/1000$, corresponding to the surface charge density, and surface current density of the ideal on-state D-checkerboard at (a) $\omega/(2\pi)=1.45\,\U{THz}$, (b) $1.76\,\U{THz}$, (c) $1.96\,\U{THz}$, and (d) $2.37\,\U{THz}$ for $y$-polarized plane waves normally incident from $z>0$ to $z<0$. The phase of $D_z$ is shifted by $90^\circ$ compared with the surface current.
}
\label{fig:d-checkerboard_on_current_and_charge_distributions}
\end{figure}
\begin{figure*}[!t]
\centering
\includegraphics[width=0.98\linewidth]{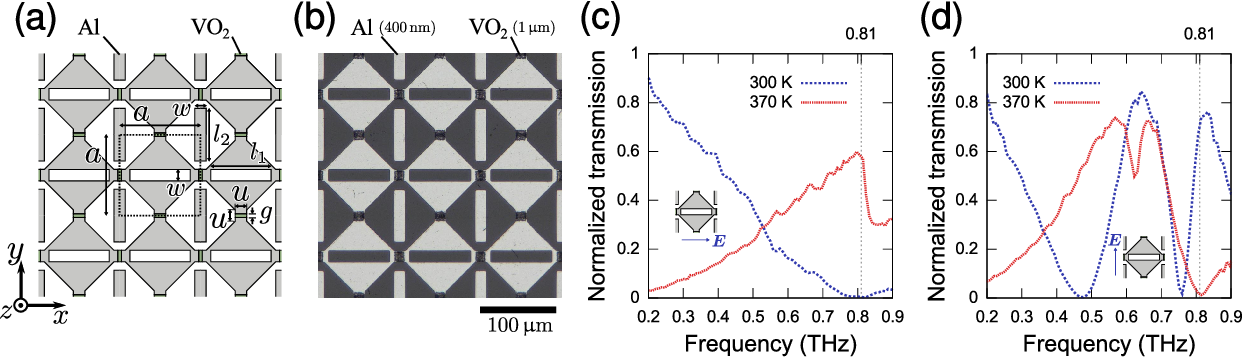}
\caption{(a) Design of a dipole-nested checkerboard metasurface on a sapphire substrate.  (b) Micrograph of the fabricated D-checkerboard on a sapphire substrate
with the target lengths $a=106\,\U{um}$, $u=w=15\,\U{um}$, $g=5\,\U{um}$, $l_1=80\,\U{um}$, and $l_2=69\,\U{um}$. The thicknesses of $\mathrm{VO}_2$ and
 Al are $1\,\U{um}$ and $400\,\U{nm}$, respectively.
 Normalized power transmission spectra of the fabricated D-checkerboard for normal incidence of a collimated terahertz beam with (c) $x$ and (d) $y$ polarization.}
\label{fig:experimantal_data}
\end{figure*}

We design the reconfigurable terahertz polarizer by using the finite
element method solver (\textsc{COMSOL Multiphysics}).
To break the 4-fold symmetry of the checkerboard, we nest dipoles with their complementary
pairs into the checkerboard so that the sufficient condition 
to realize transmission inversion for each orthogonal polarization 
can be satisfied. The design of a
dipole-nested Babinet-invertible checkerboard metasurface
(D-checkerboard) on $z=0$ is shown in
Fig.~\ref{fig:switchable_d-checkerboard_and_transmission}(a). To
simplify the analysis, we assume that all the metal sections are
zero-thickness perfect electric conductors. The calculations are
performed in the unit cell and periodic boundary conditions are applied
to the sides. The top and bottom boundaries are set as ports
with only fundamental modes. The plane wave generated by the 
top port normally enters a metasurface from $z>0$ to $z<0$. The basic
parameters are set as $a=106\,\U{um}$ and $u=w=15\,\U{um}$. The
dipoles and their complements respond only for $y$ polarization. For
$x$-polarized normally incident waves, the structure responds like a
checkerboard without the dipoles. To realize
$\tilde{t}\sur{(off)}_x(\omega_0)\sim 0$, we set a target angular
frequency of $\omega_0\sim 2\pi \times 2.4\, \U{THz}$ near the resonant
frequency of the checkerboard. Tuning the dipole length $l$ to realize
$\tilde{t}\sur{(off)}_y(\omega_0)\sim 1$, we obtain $l=67\,
\U{um}$. Figures~\ref{fig:switchable_d-checkerboard_and_transmission}(b)
and (c) show the calculated power transmission spectra of off- and
on-state D-checkerboards. We observe a 90$^\circ$ transmitting-axis
rotation around $\omega=\omega_0$. Due to the resonance nature at
$\omega=\omega_0$, an extremely high extinction ratio is achieved in
spite of the deep-subwavelength thickness of the metasurface. We also
observe that the power transmission spectra are accurately inverted with
respect to power transmission $T=1/2$ under the switching. This
phenomenon can be explained by Eq.~(\ref{eq:4}). Due
to the mirror symmetry with respect to the $x$ and $y$ axes for the
D-checkerboard, conversion between $x$ and $y$ polarizations does not
occur. Then, the power conservation law gives
$|\tilde{t}\sur{(off)}_i(\omega)|^2
+|\tilde{r}\sur{(off)}_i(\omega)|^2=1$ at the angular frequency
$\omega\leq \omega\sub{d} =2\pi c/a=2\pi \times 2.83\,\U{THz}$ without
any diffraction to higher-order modes, where $c$ is the speed of light
in a vacuum. Using Eq.~(\ref{eq:4}),
$|\tilde{t}\sur{(off)}_i(\omega)|^2
+|\tilde{r}\sur{(off)}_i(\omega)|^2=1$ can be written as
$|\tilde{t}\sur{(off)}_i(\omega)|^2
+|\tilde{t}\sur{(on)}_i(\omega)|^2=1$. Figure~\ref{fig:d-checkerboard_on_current_and_charge_distributions}
shows the charge and surface current density of the on-state
D-checkerboard at $1.45\,\U{THz}$, $1.76\,\U{THz}$, $1.96\,\U{THz}$, and
$2.37\,\U{THz}$ for $y$-polarized plane waves normally incident from $z>0$ to $z<0$. The field distributions at $1.45\,\U{THz}$ and
$2.37\,\U{THz}$ correspond to those of the on-state D-checkerboard {\it
without complementary dipoles} shown in Figs.~\ref{fig:checkerboard_on_with_dipole}(b) and (c) of the Appendix. These distributions are not greatly influenced
by the introduction of complementary dipoles. On the other hand, the
transmission dip at $1.76\,\U{THz}$ and peak at $1.96\,\U{THz}$ are
caused by hybridization of the complementary dipole mode and those of
the on-state D-checkerboards without complementary dipoles.
The further detail of multimode formation in D-checkerboard is
discussed in Appendix~\ref{app:1}.

\section{Experiment}
Now, we consider the experimental demonstration of a reconfigurable
D-checkerboard in the terahertz frequency range. For the
variable-resistance sheets, we use vanadium dioxide (VO$_2$) on a c-cut
sapphire substrate. As the temperature increases above $T\sub{c}\approx
340\,\U{K}$, VO$_2$ exhibits an insulator-to-metal transition where the
electrical conductivity typically changes by several orders of
magnitude; thus, the metasurface transitions from the off state to the
on state. Using the c-cut sapphire as a substrate at $z\leq 0$ causes
mirror symmetry breaking with respect to $z=0$ where the structures are
located, and the condition for Babinet's principle is no longer
satisfied. In the target terahertz frequency range, c-cut sapphire has a
refractive index $n_x=n_y=n_ \perp\approx 3.1$ and
$n_z=n_\parallel\approx 3.4$ \cite{Grischkowsky1990}. To compensate this
substrate effect, we must modify the design of the D-checkerboard, as
shown in Fig.~\ref{fig:experimantal_data}(a). In the following
discussion, we use the normalized transmission coefficient of the
metasurface with a sapphire substrate as $\hat{t}_i=\tilde{t}_i/t_0$ for
an $i$-polarized normally incident wave, where 
$\tilde{t}_i$ is the amplitude transmission coefficient of the plane $z=0$
with metasurface from air to the substrate,
and $t_0=2/(1+n_\perp)$ is the Fresnel coefficient for transmission 
from air to the sapphire. For $x$-polarized
normally incident waves, we changed the gap $g=5\,\U{um}$ to ensure
$\hat{t}\sur{(off)}_x(\omega_0')\sim 0$ and
$\hat{t}\sur{(on)}_x(\omega_0')\sim 1$ \cite{Urade2016}, where
$\omega_0'$ is the modified resonant angular frequency. For
$y$-polarized normally incident waves, we use different lengths for the
dipoles and their complements. By optimizing these parameters, we obtain
$l_1=80\,\U{um}$ and $l_2=69\,\U{um}$ to realize
$\hat{t}\sur{(off)}_y(\omega_0')\sim 1$ and
$\hat{t}\sur{(on)}_y(\omega_0')\sim 0$. The calculated transmission
spectra of the designed metasurface are shown in Fig.~\ref{fig:simulated_spectra_of_d-checkerboard_on_sapphire} of the Appendix.

Next, we experimentally demonstrate dynamic rotation of the
transmitting axis of the reconfigurable metasurface.
The designed D-checkerboard is fabricated as follows. A thin film of
stoichiometric VO$_2$ of about 1-$\U{um}$ thickness is deposited on a
sapphire (0001) substrate (thickness: $1.0\,\U{mm}$) via reactive
magnetron sputtering of a vanadium target \cite{Okimura2006}. After a
positive photoresist is patterned on the VO$_2$ using a maskless
lithography technique, the unnecessary part of the VO$_2$ film is
removed by wet etching. The D-checkerboard made of aluminum (thickness:
$400\,\U{nm}$) is formed by photolithography and electron beam
evaporation at room temperature and lift-off techniques.
A micrograph of the
fabricated sample is shown in Fig.~\ref{fig:experimantal_data}(b).
The $\mathrm{VO}_2$ squares
with target side length $u$ are partially overlapped by the Al structures, for ensuring electrical connection.
The maximum thickness of
the device fabricated on a c-cut sapphire substrate is $\sim
1.4\,\U{um}$, which is deep subwavelength compared with the wavelength
at the target frequency.

To evaluate the D-checkerboard, we use a conventional terahertz
time-domain spectroscopy system. To delay the signals reflected at the
boundaries, two pieces of c-cut sapphire substrates of thickness
$1\,\U{mm}$ are attached under the metasurface with a $1\,\U{mm}$ c-cut
sapphire substrate. Therefore, the overall thickness of the substrate is
$3\,\U{mm}$. A reference c-cut sapphire sample with thickness
$3\,\U{mm}$ is prepared by stacking three sapphire plates of thickness
$1\,\U{mm}$. The stacked samples are held by a brass holder with a
temperature feedback system to control the electric current in a
nichrome wire attached to the holder, while the temperature of the
holder is monitored by a thermocouple. A collimated terahertz beam with
a linear polarization and beam diameter $\sim 8\, \U{mm}$ is normally
incident onto the stacked samples. The temporal profile of the electric
field is measured using a detector dipole antenna. The multiple
reflected signal is cut using a time-domain window function.
Measuring the electric fields $E\sub{ref}(t)$
of the c-cut reference sapphire substrate and $E(t)$,
we calculate the
normalized complex amplitude transmission coefficient
$\hat{t}(\omega)=\tilde{E}(\omega)/\tilde{E}\sub{ref}(\omega)$, where
$\tilde{E}(\omega)$ and $\tilde{E}\sub{ref}(\omega)$ are Fourier
transformed electric fields.
 Figures~\ref{fig:experimantal_data}(c) and (d) show the power
transmission spectra $|\hat{t}|^2$ of the D-checkerboard on a sapphire
substrate for normal incidence of collimated terahertz beams with $x$
and $y$ polarization, respectively. The lowest diffraction frequency of
the D-checkerboard is given by $f\sub{d}=c/(n_\parallel a)\sim
0.83\,\U{THz}$. At $0.81\,\U{THz}$, the dynamic rotation of the
transmitting axis of the metasurface from the $y$ axis to the $x$ axis
is realized by heating the device. The extinction ratios
$|\hat{t}\sur{(off)}_y(\omega_0')/\hat{t}\sur{(off)}_x(\omega_0')|^2$
and $|\hat{t}\sur{(on)}_x(\omega_0')/\hat{t}\sur{(on)}_y(\omega_0')|^2$
at $\omega_0'=2\pi \times 0.81\,\U{THz}$ are $\sim 10^2$. Although the
frequency dependences of the transmission spectra agree well with the
simulated data shown in Fig.~\ref{fig:simulated_spectra_of_d-checkerboard_on_sapphire} of the Appendix, the
maximum transmission peaks of the on- and off-state metasurface are not
as high as the simulated data. This is because the resistivity of the
VO$_2$ film is not switched ideally and the aluminum has a
finite conductivity.
High quality vanadium-dioxide films
formed by other deposition techniques, such as pulse laser deposition \cite{Nag2008},
and higher conductive metals
could improve the performance of the device.

\section{Conclusion}
In this paper, we introduced the class of anisotropic
Babinet-invertible
metasurfaces enabling transmission--reflection switching for each orthogonally polarized wave, and
experimentally demonstrated their applicability to
polarization control of electromagnetic waves. The transmission inversion for each orthogonal polarization can be considered as a manifestation of an
artificially engineered insulator--metal transition of anisotropic
metasurfaces.
The concept of anisotropic Babinet-invertible metasurfaces enabling
transmission--reflection switching
for each orthogonally polarized wave
is universal, and is independent of the implementation.
Their applications are not limited to reconfigurable
terahertz polarizers, but could find a wide range of applications for
dynamic polarization control over a broad region of the electromagnetic spectrum
in which the variable-resistance materials can be used.
Ultrafast switching could be realized if photoexcitation of carriers is employed.
Thus, the anisotropic Babinet-invertible metasurfaces will pave the way
for ultra-fast polarization-selective spectroscopy.

\section*{Acknowledgments}

The authors gratefully acknowledge the contributions of T.~Nishida and T.~McArthur.
The sample fabrication was performed with the help of Kyoto University Nano
Technology Hub, as part of the ``Nanotechnology Platform Project''
sponsored by MEXT, Japan. The present research is supported by
grant from the Murata Science Foundation and 
JSPS KAKENHI Grant Number JP16K13699.
One of the authors (Y.U.) was supported by a
JSPS Research Fellowship for Young Scientists.

\appendix

 \section{Multimode formation in D-checkerboard \label{app:1}}

\begin{figure}[t]
\centering
\includegraphics[width=\linewidth]{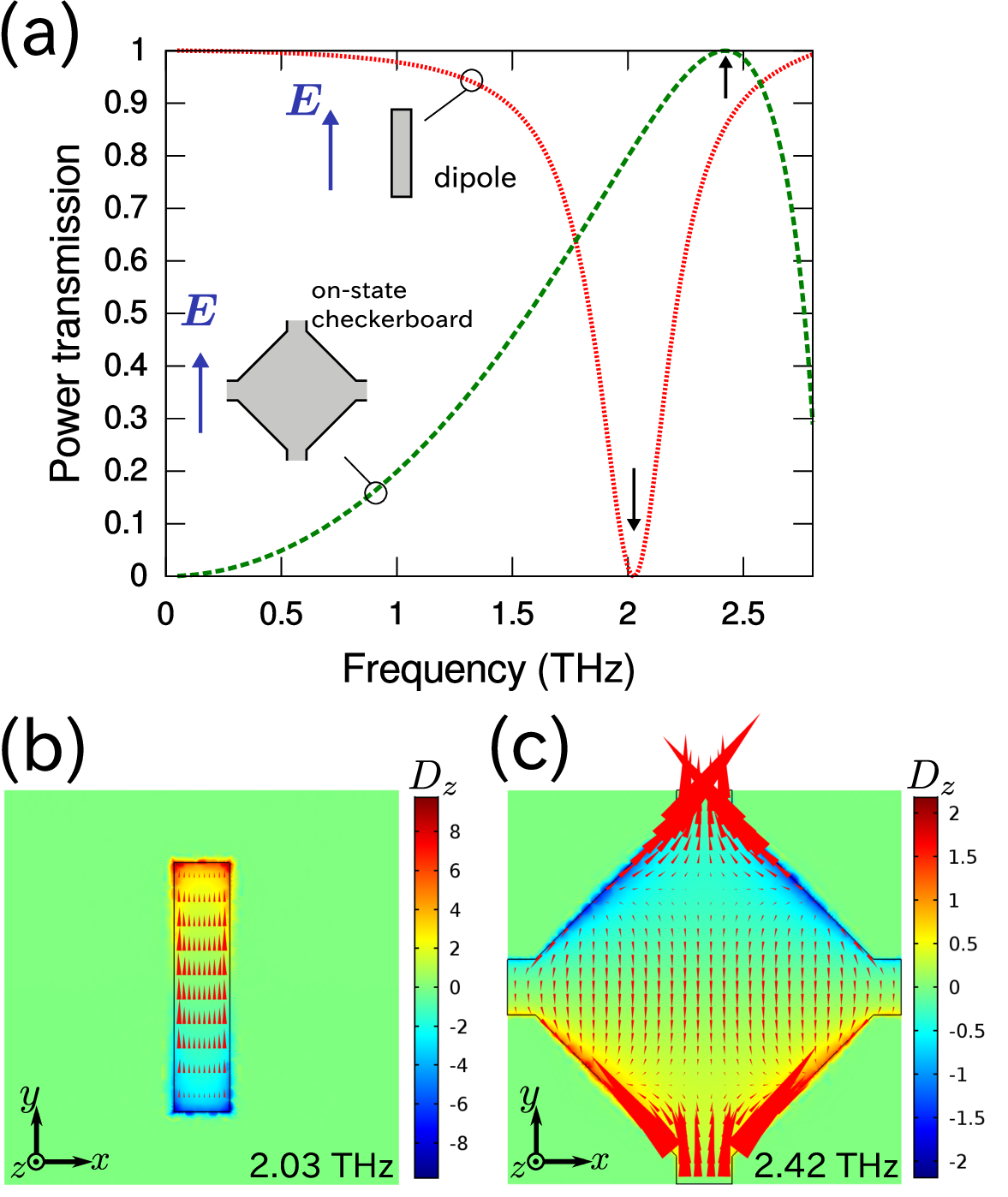}
\caption{(a) Transmission spectra of dipoles and the on-state checkerboard with $a=106\,\U{um}$, $u=w=15\,\U{um}$, and $l=67\,\U{um}$. The electric displacement $D_z$ on the plane $z=a/1000$, corresponding to the surface charge density, and surface current density of (b) dipoles at $\omega/(2\pi)=2.03\,\U{THz}$ and (c) the on-state checkerboard at $2.42\,\U{THz}$.
 The phase of $D_z$ is shifted by $90^\circ$ compared with the surface current.
}
\label{fig:dipole_and_checkerboard_on}
\end{figure}

Here, we investigate the formation of the multiple resonances of an
ideal D-checkerboard. The field distributions of on- and off-state
metasurfaces are exactly related to each other, so we focus only on the
on-state metasurface. We first study the transmission spectra and field
distribution of each component of the ideal D-checkerboard in a vacuum
by COMSOL using the parameters for an ideal D-checkerboards shown in
Fig.~2(b): $a=106\,\U{um}$, $u=w=15\,\U{um}$, and $l=67\,\U{um}$. Periodic boundary conditions are applied at the sides of the
unit cell, and ports
with only lowest frequency modes are set on
the bottom and top faces. The metallic structures are assumed to be
zero-thickness perfect electric conductors. Normally incident
$y$-polarized plane waves enter the metasurface at $z=0$ from $z>0$ to
$z<0$. Figure~\ref{fig:dipole_and_checkerboard_on}(a) shows the
transmission spectra of the dipoles and the on-state checkerboard. The
dipoles and the on-state checkerboard show a transmission dip at
$2.03\,\U{THz}$ and transmission peak at $2.42\,\U{THz}$. The electric
displacement $D_z$ just above the surface ($z=a/1000$), corresponding to
the charge density, and the surface current distributions on the metal
($z=0$) at these frequencies, are shown in
Figs.~\ref{fig:dipole_and_checkerboard_on}(b) and (c). To clearly see the
distributions, the phase of $D_z$ is shifted by $90^\circ$ compared with
the surface current in these plots. The first-order resonance of the
dipoles is shown in Fig.~\ref{fig:dipole_and_checkerboard_on}(b). In
Fig.~\ref{fig:dipole_and_checkerboard_on}(c), the direction of the
surface current at the center of the on-state checkerboard is opposite
that on the top and bottom interconnection patches.

 \begin{figure}[!t]
\centering
\includegraphics[width=\linewidth]{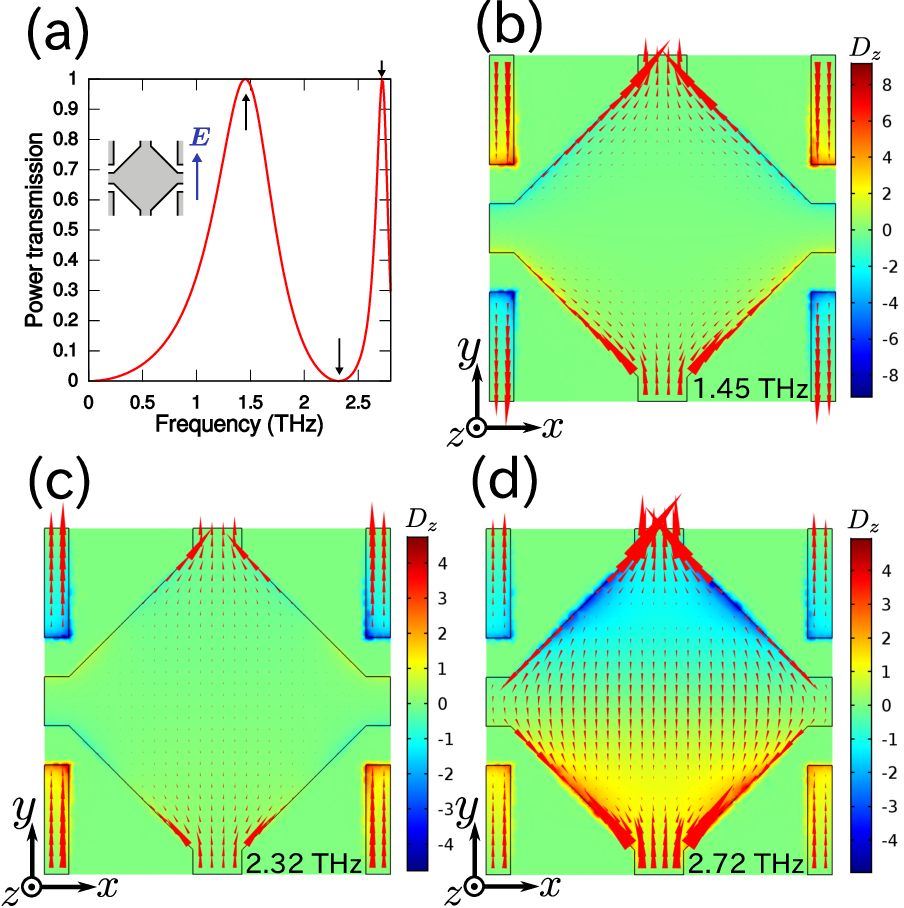}
\caption{(a) Transmission spectrum of an on-state checkerboard with dipoles having $a=106\,\U{um}$, $u=w=15\,\U{um}$, and $l=67\,\U{um}$. Electric displacement $D_z$ on the plane $z=a/1000$, corresponding to the surface charge density, and surface current density on the metasurface at (b) $\omega/(2\pi)=1.45\,\U{THz}$, (c) $2.32\,\U{THz}$, and (d) $2.72\,\U{THz}$. The phase of $D_z$ is shifted by $90^\circ$ compared with the surface current.
 }
\label{fig:checkerboard_on_with_dipole}
\end{figure}
\begin{figure}[!t]
\centering
\includegraphics[width=0.95\linewidth]{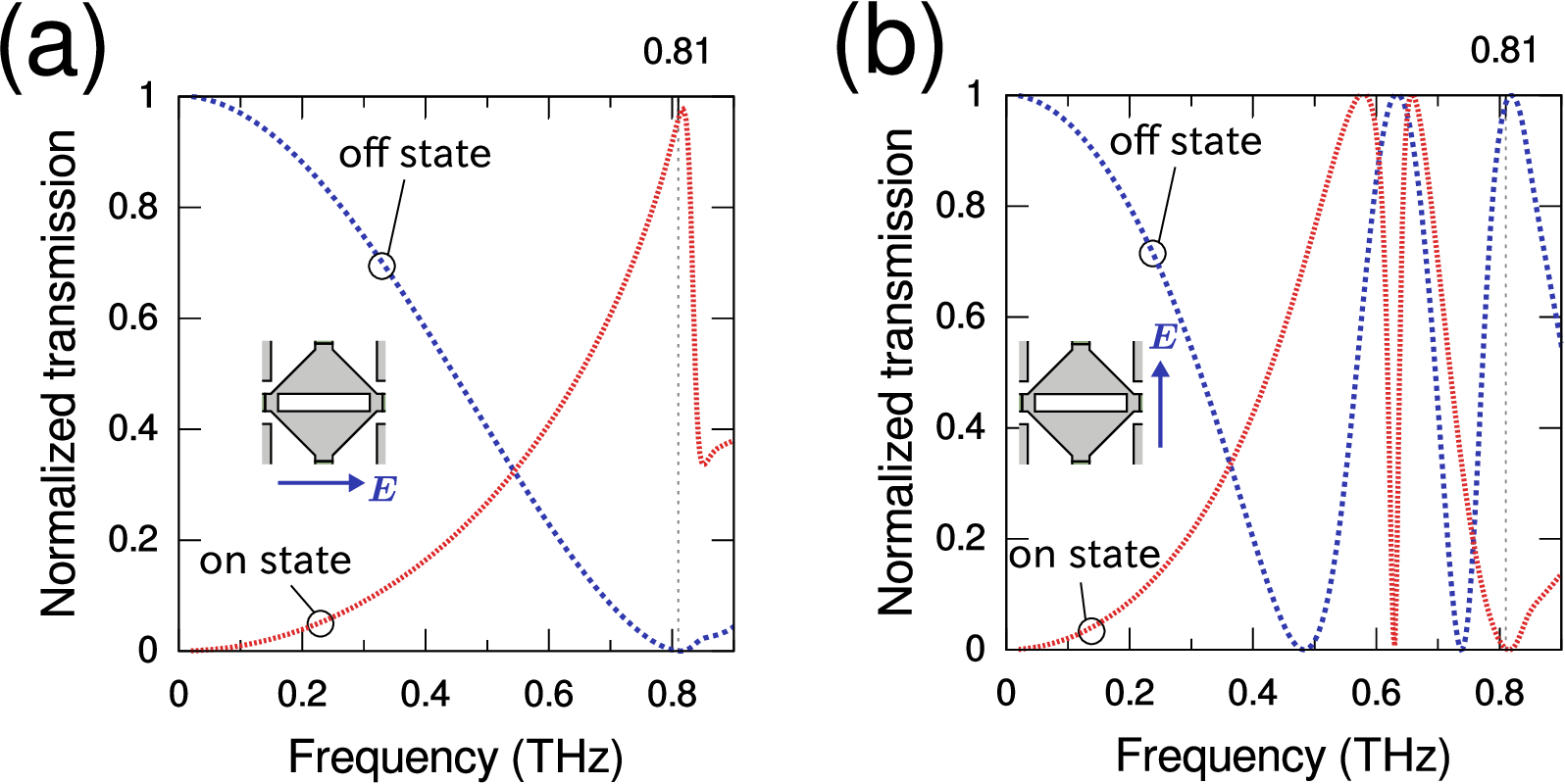}
\caption{Calculated normalized power transmission spectra of modified D-checkerboard on a c-cut sapphire substrate at the normal incidence of (a) $x$- and (b) $y$-polarized plane waves. The geometrical parameters are $a=106\,\U{um}$, $u=w=15\,\U{um}$, $g=5\,\U{um}$, $l_1=80\,\U{um}$, and $l_2=69\,\U{um}$. }
\label{fig:simulated_spectra_of_d-checkerboard_on_sapphire}
\end{figure}

Next, we consider the on-state checkerboard combined with dipoles. The
spectrum of the on-state checkerboard with dipoles at the normal
incidence of $y$-polarized plane waves is shown in
Fig.~\ref{fig:checkerboard_on_with_dipole}(a). It shows Fano-like
characteristics \cite{Miroshnichenko2010} with transmission peaks at
$1.45\,\U{THz}$ and $2.72\,\U{THz}$, and a transmission dip at
$2.32\,\U{THz}$. The charge and surface current distribution on the
dipoles at $1.45\,\U{THz}$
[Fig.~\ref{fig:checkerboard_on_with_dipole}(b)] and $2.32\,\U{THz}$
[Fig.~\ref{fig:checkerboard_on_with_dipole}(c)] are similar to those
shown in Fig.~\ref{fig:dipole_and_checkerboard_on}(b). At these
frequencies, the phase relations between the currents on the
checkerboard and the dipole are opposite, and the resonant transmission
at $1.45\,\U{THz}$ and reflection at $2.32\,\U{THz}$ are caused by
interference between the first-order dipole mode and the on-state
checkerboard mode. The current distributions on the checkerboard and the
dipole at $2.72\,\U{THz}$ in
Fig.~\ref{fig:checkerboard_on_with_dipole}(d) are similar to those in
Fig.~\ref{fig:dipole_and_checkerboard_on}(b) and (c), respectively. This
indicates that the $2.72\,\U{THz}$ resonant transmission originates
from a hybridized mode of the first-order dipole mode and
$2.42\,\U{THz}$ on-state checkerboard mode.

\section{Simulation for modified D-checkerboard on a c-cut sapphire substrate \label{app:2}}

Here, we present spectra of a modified D-checkerboard on a c-cut
sapphire substrate. The geometrical parameters are $a=106\,\U{um}$,
$u=w=15\,\U{um}$, $g=5\,\U{um}$, $l_1=80\,\U{um}$, and
$l_2=69\,\U{um}$. We assume that all conductive parts
are zero-thickness perfect
electric conductors. The calculation is performed in the unit cell and
periodic boundary conditions are imposed on the sides. The metallic
structures are located on $z=0$. A linearly-polarized plane wave is
generated from a PML (perfectly matched layer)-backed port with a slit
condition on the interior port, and detected by another port with the
similar setting. The refractive index of the c-cut sapphire is set to be
$n_x=n_y=n_ \perp= 3.1$ and $n_z=n_\parallel= 3.4$. The calculated
normalized power transmission spectra $|\hat{t}|^2 $ are shown in
Fig.~\ref{fig:simulated_spectra_of_d-checkerboard_on_sapphire}.


\begin{thebibliography}{27}%
\makeatletter
\providecommand \@ifxundefined [1]{%
 \@ifx{#1\undefined}
}%
\providecommand \@ifnum [1]{%
 \ifnum #1\expandafter \@firstoftwo
 \else \expandafter \@secondoftwo
 \fi
}%
\providecommand \@ifx [1]{%
 \ifx #1\expandafter \@firstoftwo
 \else \expandafter \@secondoftwo
 \fi
}%
\providecommand \natexlab [1]{#1}%
\providecommand \enquote  [1]{``#1''}%
\providecommand \bibnamefont  [1]{#1}%
\providecommand \bibfnamefont [1]{#1}%
\providecommand \citenamefont [1]{#1}%
\providecommand \href@noop [0]{\@secondoftwo}%
\providecommand \href [0]{\begingroup \@sanitize@url \@href}%
\providecommand \@href[1]{\@@startlink{#1}\@@href}%
\providecommand \@@href[1]{\endgroup#1\@@endlink}%
\providecommand \@sanitize@url [0]{\catcode `\\12\catcode `\$12\catcode
  `\&12\catcode `\#12\catcode `\^12\catcode `\_12\catcode `\%12\relax}%
\providecommand \@@startlink[1]{}%
\providecommand \@@endlink[0]{}%
\providecommand \url  [0]{\begingroup\@sanitize@url \@url }%
\providecommand \@url [1]{\endgroup\@href {#1}{\urlprefix }}%
\providecommand \urlprefix  [0]{URL }%
\providecommand \Eprint [0]{\href }%
\providecommand \doibase [0]{http://dx.doi.org/}%
\providecommand \selectlanguage [0]{\@gobble}%
\providecommand \bibinfo  [0]{\@secondoftwo}%
\providecommand \bibfield  [0]{\@secondoftwo}%
\providecommand \translation [1]{[#1]}%
\providecommand \BibitemOpen [0]{}%
\providecommand \bibitemStop [0]{}%
\providecommand \bibitemNoStop [0]{.\EOS\space}%
\providecommand \EOS [0]{\spacefactor3000\relax}%
\providecommand \BibitemShut  [1]{\csname bibitem#1\endcsname}%
\let\auto@bib@innerbib\@empty
\bibitem [{\citenamefont {Solymar}\ and\ \citenamefont
  {Shamonina}(2009)}]{Solymar2009}%
  \BibitemOpen
  \bibfield  {author} {\bibinfo {author} {\bibfnamefont {L.}~\bibnamefont
  {Solymar}}\ and\ \bibinfo {author} {\bibfnamefont {E.}~\bibnamefont
  {Shamonina}},\ }\href@noop {} {\emph {\bibinfo {title} {{Waves in
  Metamaterials}}}}\ (\bibinfo  {publisher} {Oxford University Press},\
  \bibinfo {address} {New York},\ \bibinfo {year} {2009})\BibitemShut {NoStop}%
\bibitem [{\citenamefont {Kildishev}\ \emph {et~al.}(2013)\citenamefont
  {Kildishev}, \citenamefont {Boltasseva},\ and\ \citenamefont
  {Shalaev}}]{Kildishev2013}%
  \BibitemOpen
  \bibfield  {author} {\bibinfo {author} {\bibfnamefont {A.~V.}\ \bibnamefont
  {Kildishev}}, \bibinfo {author} {\bibfnamefont {A.}~\bibnamefont
  {Boltasseva}}, \ and\ \bibinfo {author} {\bibfnamefont {V.~M.}\ \bibnamefont
  {Shalaev}},\ }\bibfield  {title} {\bibinfo {title} {{Planar Photonics with
  Metasurfaces}},\ }\href {\doibase 10.1126/science.1232009} {\bibfield
  {journal} {\bibinfo  {journal} {Science}\ }\textbf {\bibinfo {volume}
  {339}},\ \bibinfo {pages} {1232009} (\bibinfo {year} {2013})}\BibitemShut
  {NoStop}%
\bibitem [{\citenamefont {Monticone}\ \emph {et~al.}(2013)\citenamefont
  {Monticone}, \citenamefont {Estakhri},\ and\ \citenamefont
  {Al{\`{u}}}}]{Monticone2013}%
  \BibitemOpen
  \bibfield  {author} {\bibinfo {author} {\bibfnamefont {F.}~\bibnamefont
  {Monticone}}, \bibinfo {author} {\bibfnamefont {N.~M.}\ \bibnamefont
  {Estakhri}}, \ and\ \bibinfo {author} {\bibfnamefont {A.}~\bibnamefont
  {Al{\`{u}}}},\ }\bibfield  {title} {\bibinfo {title} {{Full Control of
  Nanoscale Optical Transmission with a Composite Metascreen}},\ }\href
  {\doibase 10.1103/PhysRevLett.110.203903} {\bibfield  {journal} {\bibinfo
  {journal} {Phys. Rev. Lett.}\ }\textbf {\bibinfo {volume} {110}},\ \bibinfo
  {pages} {203903} (\bibinfo {year} {2013})}\BibitemShut {NoStop}%
\bibitem [{\citenamefont {Pfeiffer}\ and\ \citenamefont
  {Grbic}(2013)}]{Pfeiffer2013}%
  \BibitemOpen
  \bibfield  {author} {\bibinfo {author} {\bibfnamefont {C.}~\bibnamefont
  {Pfeiffer}}\ and\ \bibinfo {author} {\bibfnamefont {A.}~\bibnamefont
  {Grbic}},\ }\bibfield  {title} {\bibinfo {title} {{Metamaterial Huygens'
  Surfaces: Tailoring Wave Fronts with Reflectionless Sheets}},\ }\href
  {\doibase 10.1103/PhysRevLett.110.197401} {\bibfield  {journal} {\bibinfo
  {journal} {Phys. Rev. Lett.}\ }\textbf {\bibinfo {volume} {110}},\ \bibinfo
  {pages} {197401} (\bibinfo {year} {2013})}\BibitemShut {NoStop}%
\bibitem [{\citenamefont {Tao}\ \emph {et~al.}(2011)\citenamefont {Tao},
  \citenamefont {Padilla}, \citenamefont {Zhang},\ and\ \citenamefont
  {Averitt}}]{Tao2011}%
  \BibitemOpen
  \bibfield  {author} {\bibinfo {author} {\bibfnamefont {H.}~\bibnamefont
  {Tao}}, \bibinfo {author} {\bibfnamefont {W.~J.}\ \bibnamefont {Padilla}},
  \bibinfo {author} {\bibfnamefont {X.}~\bibnamefont {Zhang}}, \ and\ \bibinfo
  {author} {\bibfnamefont {R.~D.}\ \bibnamefont {Averitt}},\ }\bibfield
  {title} {\bibinfo {title} {{Recent Progress in Electromagnetic Metamaterial
  Devices for Terahertz Applications}},\ }\href {\doibase
  10.1109/JSTQE.2010.2047847} {\bibfield  {journal} {\bibinfo  {journal} {IEEE
  J. Sel. Topics Quantum Electron.}\ }\textbf {\bibinfo {volume} {17}},\
  \bibinfo {pages} {92} (\bibinfo {year} {2011})}\BibitemShut {NoStop}%
\bibitem [{\citenamefont {Booker}(1946)}]{Booker1946}%
  \BibitemOpen
  \bibfield  {author} {\bibinfo {author} {\bibfnamefont {H.}~\bibnamefont
  {Booker}},\ }\bibfield  {title} {\bibinfo {title} {{SLOT AERIALS AND THEIR
  RELATION TO COMPLEMENTARY WIRE AERIALS (BABINET'S PRINCIPLE)}},\ }\href@noop
  {} {\bibfield  {journal} {\bibinfo  {journal} {J. IEE (London), part IIIA}\
  }\textbf {\bibinfo {volume} {93}},\ \bibinfo {pages} {620} (\bibinfo {year}
  {1946})}\BibitemShut {NoStop}%
\bibitem [{Note1()}]{Note1}%
  \BibitemOpen
  \bibinfo {note} {Rigorously, the sheet impedance is defined as $\protect
  \mathaccentV {tilde}07E{\protect \mathbf {E}}_\parallel (x,y)=Z_\protect
  \textnormal {s}(x,y) \protect \mathaccentV {tilde}07E{\protect \mathbf
  {K}}(x,y)$, for the tangential component of the electric field $\protect
  \mathaccentV {tilde}07E{\protect \mathbf {E}}_\parallel (x,y)\protect
  \qopname \relax o{exp}(-{\protect \mathrm {i}}\omega t)+\protect \textnormal
  {c.c.}$ on $z=0$ and surface current density $\protect \mathaccentV
  {tilde}07E{\protect \mathbf {K}}(x,y)\protect \qopname \relax
  o{exp}(-{\protect \mathrm {i}}\omega t)+\protect \textnormal {c.c.}$ on
  $z=0$. Here, $\omega $ is an angular frequency and ``$\protect \textnormal
  {c.c.}$'' represents the complex conjugate operation.}\BibitemShut {Stop}%
\bibitem [{\citenamefont {Baum}\ and\ \citenamefont
  {Singaraju}(1974)}]{Baum1974}%
  \BibitemOpen
  \bibfield  {author} {\bibinfo {author} {\bibfnamefont {C.~E.}\ \bibnamefont
  {Baum}}\ and\ \bibinfo {author} {\bibfnamefont {B.~K.}\ \bibnamefont
  {Singaraju}},\ }\bibfield  {title} {\bibinfo {title} {{Generalization of
  Babinet's Principle in Terms of the Combined Field to Include Impedance
  Loaded Aperture Antennas and Scatterers}},\ }\href@noop {} {\bibfield
  {journal} {\bibinfo  {journal} {Interaction Note No.217 (Air Force Weapons
  Lab., Kirtland Air Force Base, NM 87117)}\ } (\bibinfo {year}
  {1974})}\BibitemShut {NoStop}%
\bibitem [{\citenamefont {Compton}\ \emph {et~al.}(1984)\citenamefont
  {Compton}, \citenamefont {Macfarlane}, \citenamefont {Whitbourn},
  \citenamefont {Blanco},\ and\ \citenamefont {McPhedran}}]{Compton1984}%
  \BibitemOpen
  \bibfield  {author} {\bibinfo {author} {\bibfnamefont {R.~C.}\ \bibnamefont
  {Compton}}, \bibinfo {author} {\bibfnamefont {J.~C.}\ \bibnamefont
  {Macfarlane}}, \bibinfo {author} {\bibfnamefont {L.~B.}\ \bibnamefont
  {Whitbourn}}, \bibinfo {author} {\bibfnamefont {M.~M.}\ \bibnamefont
  {Blanco}}, \ and\ \bibinfo {author} {\bibfnamefont {R.~C.}\ \bibnamefont
  {McPhedran}},\ }\bibfield  {title} {\bibinfo {title} {{Babinet's principle
  applied to ideal beam-splitters for submillimetre waves}},\ }\href {\doibase
  10.1080/713821538} {\bibfield  {journal} {\bibinfo  {journal} {Opt. Acta}\
  }\textbf {\bibinfo {volume} {31}},\ \bibinfo {pages} {515} (\bibinfo {year}
  {1984})}\BibitemShut {NoStop}%
\bibitem [{\citenamefont {Kempa}(2010)}]{Kempa2010}%
  \BibitemOpen
  \bibfield  {author} {\bibinfo {author} {\bibfnamefont {K.}~\bibnamefont
  {Kempa}},\ }\bibfield  {title} {\bibinfo {title} {{Percolation effects in the
  checkerboard Babinet series of metamaterial structures}},\ }\href {\doibase
  10.1002/pssr.201004266} {\bibfield  {journal} {\bibinfo  {journal} {Phys.
  Status Solidi Rapid Res. Lett.}\ }\textbf {\bibinfo {volume} {4}},\ \bibinfo
  {pages} {218} (\bibinfo {year} {2010})}\BibitemShut {NoStop}%
\bibitem [{\citenamefont {Edmunds}\ \emph {et~al.}(2010)\citenamefont
  {Edmunds}, \citenamefont {Hibbins}, \citenamefont {Sambles},\ and\
  \citenamefont {Youngs}}]{Edmunds2010}%
  \BibitemOpen
  \bibfield  {author} {\bibinfo {author} {\bibfnamefont {J.~D.}\ \bibnamefont
  {Edmunds}}, \bibinfo {author} {\bibfnamefont {A.~P.}\ \bibnamefont
  {Hibbins}}, \bibinfo {author} {\bibfnamefont {J.~R.}\ \bibnamefont
  {Sambles}}, \ and\ \bibinfo {author} {\bibfnamefont {I.~J.}\ \bibnamefont
  {Youngs}},\ }\bibfield  {title} {\bibinfo {title} {{Resonantly inverted
  microwave transmissivity threshold of metal grids}},\ }\href {\doibase
  10.1088/1367-2630/12/6/063007} {\bibfield  {journal} {\bibinfo  {journal}
  {New J. Phys.}\ }\textbf {\bibinfo {volume} {12}},\ \bibinfo {pages} {063007}
  (\bibinfo {year} {2010})}\BibitemShut {NoStop}%
\bibitem [{\citenamefont {Takano}\ \emph {et~al.}(2014)\citenamefont {Takano},
  \citenamefont {Miyamaru}, \citenamefont {Akiyama}, \citenamefont {Miyazaki},
  \citenamefont {Takeda}, \citenamefont {Abe}, \citenamefont {Tokuda},
  \citenamefont {Ito},\ and\ \citenamefont {Hangyo}}]{Takano2014}%
  \BibitemOpen
  \bibfield  {author} {\bibinfo {author} {\bibfnamefont {K.}~\bibnamefont
  {Takano}}, \bibinfo {author} {\bibfnamefont {F.}~\bibnamefont {Miyamaru}},
  \bibinfo {author} {\bibfnamefont {K.}~\bibnamefont {Akiyama}}, \bibinfo
  {author} {\bibfnamefont {H.}~\bibnamefont {Miyazaki}}, \bibinfo {author}
  {\bibfnamefont {M.~W.}\ \bibnamefont {Takeda}}, \bibinfo {author}
  {\bibfnamefont {Y.}~\bibnamefont {Abe}}, \bibinfo {author} {\bibfnamefont
  {Y.}~\bibnamefont {Tokuda}}, \bibinfo {author} {\bibfnamefont
  {H.}~\bibnamefont {Ito}}, \ and\ \bibinfo {author} {\bibfnamefont
  {M.}~\bibnamefont {Hangyo}},\ }\bibfield  {title} {\bibinfo {title}
  {{Crossover from capacitive to inductive electromagnetic responses in near
  self-complementary metallic checkerboard patterns}},\ }\href {\doibase
  10.1364/OE.22.024787} {\bibfield  {journal} {\bibinfo  {journal} {Opt.
  Express}\ }\textbf {\bibinfo {volume} {22}},\ \bibinfo {pages} {24787}
  (\bibinfo {year} {2014})}\BibitemShut {NoStop}%
\bibitem [{\citenamefont {Ramakrishna}\ \emph {et~al.}(2011)\citenamefont
  {Ramakrishna}, \citenamefont {Mandal}, \citenamefont {Jeyadheepan},
  \citenamefont {Shukla}, \citenamefont {Chakrabarti}, \citenamefont {Kadic},
  \citenamefont {Enoch},\ and\ \citenamefont {Guenneau}}]{Ramakrishna2011}%
  \BibitemOpen
  \bibfield  {author} {\bibinfo {author} {\bibfnamefont {S.~A.}\ \bibnamefont
  {Ramakrishna}}, \bibinfo {author} {\bibfnamefont {P.}~\bibnamefont {Mandal}},
  \bibinfo {author} {\bibfnamefont {K.}~\bibnamefont {Jeyadheepan}}, \bibinfo
  {author} {\bibfnamefont {N.}~\bibnamefont {Shukla}}, \bibinfo {author}
  {\bibfnamefont {S.}~\bibnamefont {Chakrabarti}}, \bibinfo {author}
  {\bibfnamefont {M.}~\bibnamefont {Kadic}}, \bibinfo {author} {\bibfnamefont
  {S.}~\bibnamefont {Enoch}}, \ and\ \bibinfo {author} {\bibfnamefont
  {S.}~\bibnamefont {Guenneau}},\ }\bibfield  {title} {\bibinfo {title}
  {{Plasmonic interaction of visible light with gold nanoscale
  checkerboards}},\ }\href {\doibase 10.1103/PhysRevB.84.245424} {\bibfield
  {journal} {\bibinfo  {journal} {Phys. Rev. B}\ }\textbf {\bibinfo {volume}
  {84}},\ \bibinfo {pages} {245424} (\bibinfo {year} {2011})}\BibitemShut
  {NoStop}%
\bibitem [{\citenamefont {Gonz{\'{a}}lez-Ovejero}\ \emph
  {et~al.}(2015{\natexlab{a}})\citenamefont {Gonz{\'{a}}lez-Ovejero},
  \citenamefont {Martini}, \citenamefont {Loiseaux}, \citenamefont
  {Tripon-Canseliet}, \citenamefont {Mencagli}, \citenamefont {Chazelas},\ and\
  \citenamefont {Maci}}]{Gonzalez-Ovejero2014a}%
  \BibitemOpen
  \bibfield  {author} {\bibinfo {author} {\bibfnamefont {D.}~\bibnamefont
  {Gonz{\'{a}}lez-Ovejero}}, \bibinfo {author} {\bibfnamefont {E.}~\bibnamefont
  {Martini}}, \bibinfo {author} {\bibfnamefont {B.}~\bibnamefont {Loiseaux}},
  \bibinfo {author} {\bibfnamefont {C.}~\bibnamefont {Tripon-Canseliet}},
  \bibinfo {author} {\bibfnamefont {M.~J.}\ \bibnamefont {Mencagli}}, \bibinfo
  {author} {\bibfnamefont {J.}~\bibnamefont {Chazelas}}, \ and\ \bibinfo
  {author} {\bibfnamefont {S.}~\bibnamefont {Maci}},\ }\bibfield  {title}
  {\bibinfo {title} {{Basic Properties of Checkerboard Metasurfaces}},\ }\href
  {\doibase 10.1109/LAWP.2014.2365021} {\bibfield  {journal} {\bibinfo
  {journal} {IEEE Antennas Wireless Propag. Lett.}\ }\textbf {\bibinfo {volume}
  {14}},\ \bibinfo {pages} {406} (\bibinfo {year}
  {2015}{\natexlab{a}})}\BibitemShut {NoStop}%
\bibitem [{\citenamefont {Gonz{\'{a}}lez-Ovejero}\ \emph
  {et~al.}(2015{\natexlab{b}})\citenamefont {Gonz{\'{a}}lez-Ovejero},
  \citenamefont {Martini},\ and\ \citenamefont {Maci}}]{Gonzalez-Ovejero2014}%
  \BibitemOpen
  \bibfield  {author} {\bibinfo {author} {\bibfnamefont {D.}~\bibnamefont
  {Gonz{\'{a}}lez-Ovejero}}, \bibinfo {author} {\bibfnamefont {E.}~\bibnamefont
  {Martini}}, \ and\ \bibinfo {author} {\bibfnamefont {S.}~\bibnamefont
  {Maci}},\ }\bibfield  {title} {\bibinfo {title} {{Surface Waves Supported by
  Metasurfaces With Self-Complementary Geometries}},\ }\href {\doibase
  10.1109/TAP.2014.2367535} {\bibfield  {journal} {\bibinfo  {journal} {IEEE
  Trans. Antennas Propag.}\ }\textbf {\bibinfo {volume} {63}},\ \bibinfo
  {pages} {250} (\bibinfo {year} {2015}{\natexlab{b}})}\BibitemShut {NoStop}%
\bibitem [{\citenamefont {Tremain}\ \emph {et~al.}(2015)\citenamefont
  {Tremain}, \citenamefont {Durrant}, \citenamefont {Carter}, \citenamefont
  {Hibbins},\ and\ \citenamefont {Sambles}}]{Tremain2015}%
  \BibitemOpen
  \bibfield  {author} {\bibinfo {author} {\bibfnamefont {B.}~\bibnamefont
  {Tremain}}, \bibinfo {author} {\bibfnamefont {C.~J.}\ \bibnamefont
  {Durrant}}, \bibinfo {author} {\bibfnamefont {I.~E.}\ \bibnamefont {Carter}},
  \bibinfo {author} {\bibfnamefont {A.~P.}\ \bibnamefont {Hibbins}}, \ and\
  \bibinfo {author} {\bibfnamefont {J.~R.}\ \bibnamefont {Sambles}},\
  }\bibfield  {title} {\bibinfo {title} {{The Effect of Rotational Disorder on
  the Microwave Transmission of Checkerboard Metal Square Arrays}},\ }\href
  {\doibase 10.1038/srep16608} {\bibfield  {journal} {\bibinfo  {journal} {Sci.
  Rep.}\ }\textbf {\bibinfo {volume} {5}},\ \bibinfo {pages} {16608} (\bibinfo
  {year} {2015})}\BibitemShut {NoStop}%
\bibitem [{\citenamefont {Nakata}\ \emph {et~al.}(2013)\citenamefont {Nakata},
  \citenamefont {Urade}, \citenamefont {Nakanishi},\ and\ \citenamefont
  {Kitano}}]{Nakata2013}%
  \BibitemOpen
  \bibfield  {author} {\bibinfo {author} {\bibfnamefont {Y.}~\bibnamefont
  {Nakata}}, \bibinfo {author} {\bibfnamefont {Y.}~\bibnamefont {Urade}},
  \bibinfo {author} {\bibfnamefont {T.}~\bibnamefont {Nakanishi}}, \ and\
  \bibinfo {author} {\bibfnamefont {M.}~\bibnamefont {Kitano}},\ }\bibfield
  {title} {\bibinfo {title} {{Plane-wave scattering by self-complementary
  metasurfaces in terms of electromagnetic duality and Babinet's principle}},\
  }\href {\doibase 10.1103/PhysRevB.88.205138} {\bibfield  {journal} {\bibinfo
  {journal} {Phys. Rev. B}\ }\textbf {\bibinfo {volume} {88}},\ \bibinfo
  {pages} {205138} (\bibinfo {year} {2013})}\BibitemShut {NoStop}%
\bibitem [{\citenamefont {Urade}\ \emph {et~al.}(2015)\citenamefont {Urade},
  \citenamefont {Nakata}, \citenamefont {Nakanishi},\ and\ \citenamefont
  {Kitano}}]{Urade2015}%
  \BibitemOpen
  \bibfield  {author} {\bibinfo {author} {\bibfnamefont {Y.}~\bibnamefont
  {Urade}}, \bibinfo {author} {\bibfnamefont {Y.}~\bibnamefont {Nakata}},
  \bibinfo {author} {\bibfnamefont {T.}~\bibnamefont {Nakanishi}}, \ and\
  \bibinfo {author} {\bibfnamefont {M.}~\bibnamefont {Kitano}},\ }\bibfield
  {title} {\bibinfo {title} {{Frequency-Independent Response of
  Self-Complementary Checkerboard Screens}},\ }\href {\doibase
  10.1103/PhysRevLett.114.237401} {\bibfield  {journal} {\bibinfo  {journal}
  {Phys. Rev. Lett.}\ }\textbf {\bibinfo {volume} {114}},\ \bibinfo {pages}
  {237401} (\bibinfo {year} {2015})}\BibitemShut {NoStop}%
\bibitem [{\citenamefont {Urade}\ \emph
  {et~al.}(2016{\natexlab{a}})\citenamefont {Urade}, \citenamefont {Nakata},
  \citenamefont {Nakanishi},\ and\ \citenamefont {Kitano}}]{Urade2016a}%
  \BibitemOpen
  \bibfield  {author} {\bibinfo {author} {\bibfnamefont {Y.}~\bibnamefont
  {Urade}}, \bibinfo {author} {\bibfnamefont {Y.}~\bibnamefont {Nakata}},
  \bibinfo {author} {\bibfnamefont {T.}~\bibnamefont {Nakanishi}}, \ and\
  \bibinfo {author} {\bibfnamefont {M.}~\bibnamefont {Kitano}},\ }\bibfield
  {title} {\bibinfo {title} {{Broadband and energy-concentrating terahertz
  coherent perfect absorber based on a self-complementary metasurface}},\
  }\href {\doibase 10.1364/OL.41.004472} {\bibfield  {journal} {\bibinfo
  {journal} {Opt. Lett.}\ }\textbf {\bibinfo {volume} {41}},\ \bibinfo {pages}
  {4472} (\bibinfo {year} {2016}{\natexlab{a}})}\BibitemShut {NoStop}%
\bibitem [{Note2()}]{Note2}%
  \BibitemOpen
  \bibinfo {note} {Note that the complex amplitude transmission coefficient
  $\protect \mathaccentV {tilde}07E{t}$ from an incident wave $\protect
  \mathaccentV {check}014{\protect \mathbf {E}}_\protect \textnormal
  {0}\protect \qopname \relax o{exp}[{\protect \mathrm {i}}(\protect \mathbf
  {k}_\protect \textnormal {0}\cdot \protect \mathbf {x}-\omega t)]+\protect
  \textnormal {c.c.}$ to a transmitted wave $\protect \mathaccentV
  {check}014{\protect \mathbf {E}}_\protect \textnormal {t}\protect \qopname
  \relax o{exp}[{\protect \mathrm {i}}(\protect \mathbf {k}_\protect
  \textnormal {t}\cdot \protect \mathbf {x}-\omega t)]+\protect \textnormal
  {c.c.}$ is defined by $\protect \mathaccentV {check}014{\protect \mathbf
  {E}}_\protect \textnormal {t}=\protect \mathaccentV {tilde}07E{t}\protect
  \mathaccentV {check}014{\protect \mathbf {E}}_\protect \textnormal {0}$. In
  this paper, the modifier {\protect \it zeroth-order} represents the case that
  $\protect \mathbf {k}_\protect \textnormal {t}$ satisfies $\protect \mathcal
  {P} \protect \mathbf {k}_\protect \textnormal {t}=\protect \mathcal {P}
  \protect \mathbf {k}_\protect \textnormal {0}$, where $\protect \mathcal {P}:
  (v_x,v_y,v_z)^\protect \mathrm {T}\DOTSB \mapstochar \rightarrow
  (v_x,v_y,0)^\protect \mathrm {T}$ is the projection operator onto $z=0$
  ($\protect \mathrm {T}$: transpose).}\BibitemShut {Stop}%
\bibitem [{\citenamefont {Urade}\ \emph
  {et~al.}(2016{\natexlab{b}})\citenamefont {Urade}, \citenamefont {Nakata},
  \citenamefont {Okimura}, \citenamefont {Nakanishi}, \citenamefont {Miyamaru},
  \citenamefont {Takeda},\ and\ \citenamefont {Kitano}}]{Urade2016}%
  \BibitemOpen
  \bibfield  {author} {\bibinfo {author} {\bibfnamefont {Y.}~\bibnamefont
  {Urade}}, \bibinfo {author} {\bibfnamefont {Y.}~\bibnamefont {Nakata}},
  \bibinfo {author} {\bibfnamefont {K.}~\bibnamefont {Okimura}}, \bibinfo
  {author} {\bibfnamefont {T.}~\bibnamefont {Nakanishi}}, \bibinfo {author}
  {\bibfnamefont {F.}~\bibnamefont {Miyamaru}}, \bibinfo {author}
  {\bibfnamefont {M.~W.}\ \bibnamefont {Takeda}}, \ and\ \bibinfo {author}
  {\bibfnamefont {M.}~\bibnamefont {Kitano}},\ }\bibfield  {title} {\bibinfo
  {title} {{Dynamically Babinet-invertible metasurface: a capacitive-inductive
  reconfigurable filter for terahertz waves using vanadium-dioxide
  metal-insulator transition}},\ }\href {\doibase 10.1364/OE.24.004405}
  {\bibfield  {journal} {\bibinfo  {journal} {Opt. Express}\ }\textbf {\bibinfo
  {volume} {24}},\ \bibinfo {pages} {4405} (\bibinfo {year}
  {2016}{\natexlab{b}})}\BibitemShut {NoStop}%
\bibitem [{\citenamefont {van~der Valk}\ \emph {et~al.}(2005)\citenamefont
  {van~der Valk}, \citenamefont {van~der Marel},\ and\ \citenamefont
  {Planken}}]{VanderValk2005}%
  \BibitemOpen
  \bibfield  {author} {\bibinfo {author} {\bibfnamefont {N.~C.~J.}\
  \bibnamefont {van~der Valk}}, \bibinfo {author} {\bibfnamefont {W.~A.~M.}\
  \bibnamefont {van~der Marel}}, \ and\ \bibinfo {author} {\bibfnamefont
  {P.~C.~M.}\ \bibnamefont {Planken}},\ }\bibfield  {title} {\bibinfo {title}
  {{Terahertz polarization imaging}},\ }\href {\doibase 10.1364/OL.30.002802}
  {\bibfield  {journal} {\bibinfo  {journal} {Opt. Lett.}\ }\textbf {\bibinfo
  {volume} {30}},\ \bibinfo {pages} {2802} (\bibinfo {year}
  {2005})}\BibitemShut {NoStop}%
\bibitem [{\citenamefont {Katletz}\ \emph {et~al.}(2012)\citenamefont
  {Katletz}, \citenamefont {Pfleger}, \citenamefont {P{\"{u}}hringer},
  \citenamefont {Mikulics}, \citenamefont {Vieweg}, \citenamefont {Peters},
  \citenamefont {Scherger}, \citenamefont {Scheller}, \citenamefont {Koch},\
  and\ \citenamefont {Wiesauer}}]{Katletz2012}%
  \BibitemOpen
  \bibfield  {author} {\bibinfo {author} {\bibfnamefont {S.}~\bibnamefont
  {Katletz}}, \bibinfo {author} {\bibfnamefont {M.}~\bibnamefont {Pfleger}},
  \bibinfo {author} {\bibfnamefont {H.}~\bibnamefont {P{\"{u}}hringer}},
  \bibinfo {author} {\bibfnamefont {M.}~\bibnamefont {Mikulics}}, \bibinfo
  {author} {\bibfnamefont {N.}~\bibnamefont {Vieweg}}, \bibinfo {author}
  {\bibfnamefont {O.}~\bibnamefont {Peters}}, \bibinfo {author} {\bibfnamefont
  {B.}~\bibnamefont {Scherger}}, \bibinfo {author} {\bibfnamefont
  {M.}~\bibnamefont {Scheller}}, \bibinfo {author} {\bibfnamefont
  {M.}~\bibnamefont {Koch}}, \ and\ \bibinfo {author} {\bibfnamefont
  {K.}~\bibnamefont {Wiesauer}},\ }\bibfield  {title} {\bibinfo {title}
  {{Polarization sensitive terahertz imaging: detection of birefringence and
  optical axis}},\ }\href {\doibase 10.1364/OE.20.023025} {\bibfield  {journal}
  {\bibinfo  {journal} {Opt. Express}\ }\textbf {\bibinfo {volume} {20}},\
  \bibinfo {pages} {23025} (\bibinfo {year} {2012})}\BibitemShut {NoStop}%
\bibitem [{\citenamefont {Grischkowsky}\ \emph {et~al.}(1990)\citenamefont
  {Grischkowsky}, \citenamefont {Keiding}, \citenamefont {van Exter},\ and\
  \citenamefont {Fattinger}}]{Grischkowsky1990}%
  \BibitemOpen
  \bibfield  {author} {\bibinfo {author} {\bibfnamefont {D.}~\bibnamefont
  {Grischkowsky}}, \bibinfo {author} {\bibfnamefont {S.}~\bibnamefont
  {Keiding}}, \bibinfo {author} {\bibfnamefont {M.}~\bibnamefont {van Exter}},
  \ and\ \bibinfo {author} {\bibfnamefont {C.}~\bibnamefont {Fattinger}},\
  }\bibfield  {title} {\bibinfo {title} {{Far-infrared time-domain spectroscopy
  with terahertz beams of dielectrics and semiconductors}},\ }\href {\doibase
  10.1364/JOSAB.7.002006} {\bibfield  {journal} {\bibinfo  {journal} {J. Opt.
  Soc. Am. B}\ }\textbf {\bibinfo {volume} {7}},\ \bibinfo {pages} {2006}
  (\bibinfo {year} {1990})}\BibitemShut {NoStop}%
\bibitem [{\citenamefont {Okimura}\ \emph {et~al.}(2006)\citenamefont
  {Okimura}, \citenamefont {Sasakawa},\ and\ \citenamefont
  {Nihei}}]{Okimura2006}%
  \BibitemOpen
  \bibfield  {author} {\bibinfo {author} {\bibfnamefont {K.}~\bibnamefont
  {Okimura}}, \bibinfo {author} {\bibfnamefont {Y.}~\bibnamefont {Sasakawa}}, \
  and\ \bibinfo {author} {\bibfnamefont {Y.}~\bibnamefont {Nihei}},\ }\bibfield
   {title} {\bibinfo {title} {{X-ray Diffraction Study of Electric
  Field-Induced Metal–Insulator Transition of Vanadium Dioxide Film on
  Sapphire Substrate}},\ }\href {\doibase 10.1143/JJAP.45.9200} {\bibfield
  {journal} {\bibinfo  {journal} {Jpn. J. Appl. Phys.}\ }\textbf {\bibinfo
  {volume} {45}},\ \bibinfo {pages} {9200} (\bibinfo {year}
  {2006})}\BibitemShut {NoStop}%
\bibitem [{\citenamefont {Nag}\ and\ \citenamefont {{Haglund
  Jr}}(2008)}]{Nag2008}%
  \BibitemOpen
  \bibfield  {author} {\bibinfo {author} {\bibfnamefont {J.}~\bibnamefont
  {Nag}}\ and\ \bibinfo {author} {\bibfnamefont {R.~F.}\ \bibnamefont {{Haglund
  Jr}}},\ }\bibfield  {title} {\bibinfo {title} {{Synthesis of vanadium dioxide
  thin films and nanoparticles}},\ }\href {\doibase
  10.1088/0953-8984/20/26/264016} {\bibfield  {journal} {\bibinfo  {journal}
  {J. Phys.: Condens. Matter}\ }\textbf {\bibinfo {volume} {20}},\ \bibinfo
  {pages} {264016} (\bibinfo {year} {2008})}\BibitemShut {NoStop}%
\bibitem [{\citenamefont {Miroshnichenko}\ \emph {et~al.}(2010)\citenamefont
  {Miroshnichenko}, \citenamefont {Flach},\ and\ \citenamefont
  {Kivshar}}]{Miroshnichenko2010}%
  \BibitemOpen
  \bibfield  {author} {\bibinfo {author} {\bibfnamefont {A.~E.}\ \bibnamefont
  {Miroshnichenko}}, \bibinfo {author} {\bibfnamefont {S.}~\bibnamefont
  {Flach}}, \ and\ \bibinfo {author} {\bibfnamefont {Y.~S.}\ \bibnamefont
  {Kivshar}},\ }\bibfield  {title} {\bibinfo {title} {{Fano resonances in
  nanoscale structures}},\ }\href {\doibase 10.1103/RevModPhys.82.2257}
  {\bibfield  {journal} {\bibinfo  {journal} {Rev. Mod. Phys.}\ }\textbf
  {\bibinfo {volume} {82}},\ \bibinfo {pages} {2257} (\bibinfo {year}
  {2010})}\BibitemShut {NoStop}%
\end{thebibliography}

%

\end{document}